\documentclass[aps,prb,tightenlines,showpacs,twocolumn,superscriptaddress]{revtex4-1}

\usepackage{amsmath,amssymb,amsfonts,bm}
\usepackage{graphicx}
\usepackage{dcolumn}
\usepackage[colorlinks=true,linkcolor=blue]{hyperref}%

\begin{document}


\title{The Kondo Temperature of a Two-dimensional Electron Gas with Rashba Spin-orbit Coupling}

\author{Liang Chen}
\affiliation{Beijing Computational Science Research Centre, Beijing, 100084, China}
\affiliation{Mathematics and Physics Department, North China Electric Power University, Beijing, 102206, China}
\email[Corresponding Email: ]{csl@csrc.ac.cn}
\author{Jinhua Sun}
\affiliation{Department of Physics, Zhejiang University, Hangzhou, 310058, China}
\author{Ho-Kin Tang}
\affiliation{Department of Physics, National University of Singapore, 117542, Singapore}
\author{Hai-Qing Lin}
\affiliation{Beijing Computational Science Research Centre, Beijing, 100084, China}
\date{\today}

\begin{abstract}
  We use the Hirsch-Fye quantum Monte Carlo method to study the single magnetic impurity problem in a two-dimensional electron gas with Rashba spin-orbit coupling. We calculate the spin susceptibility for various values of spin-orbit coupling, Hubbard interaction, and chemical potential. The Kondo temperatures for different parameters are estimated by  fitting the universal curves of spin susceptibility. We find that the Kondo temperature is almost a linear function of Rashba spin-orbit energy when the chemical potential is close to the edge of the conduction band. When the chemical potential is far away from the band edge, the Kondo temperature is independent of the spin-orbit coupling. These results demonstrate that, for single impurity problem in this system, the  most important reason to change the Kondo temperature is the divergence of density of states near the band edge, and the divergence is induced by the Rashba spin-orbit coupling. 
\end{abstract}


\maketitle




Spin-orbit coupling\cite{RolandWinkler2003}, the interaction between a quantum particle's spin and its motion, has attracted much attention in condensed matter physics. The spin-orbit coupling is crucial for quantum spin Hall effect\cite{PhysRevLett.95.226801,PhysRevLett.95.146802} and topological insulators\cite{RevModPhys.82.3045,RevModPhys.83.1057,Nature.464.194}. Magnetic doping in these systems is very important. It is reported that magnetic doping is an efficient method to tune and to control the transport properties of these systems. Magnetic doping is a feasible method to break the time reversal symmetry on the surface of topological insulators\cite{Chen06082010,Wray07322011,PhysRevLett.106.206805,PhysRevLett.108.117601,PhysRevB.81.195203,PhysRevLett.108.256810}, so that one may observe very interesting physical phenomena, i.e., the topological magnetoelectric effect\cite{PhysRevB.78.195424}, the half-integer quantum Hall effect\cite{PhysRevLett.106.166802}, the image magnetic monopole induced by an electric charge near the surface of topological insulator\cite{Qi27022009,PhysRevB.81.245125}, the topologically quantized magneto-optical Kerr and Faraday\cite{PhysRevB.84.205327,PhysRevLett.105.057401} rotation in units of the fine structure constant\cite{PhysRevLett.105.166803,PhysRevB.83.205109}, the repulsive Casimir interaction between topological insulators with opposite topological magnetoelectric polarizabilities\cite{PhysRevLett.106.020403,PhysRevB.84.045119,PhysRevB.84.075149}, etc. It is also demonstrated experimentally and theoretically that magnetic doping is a straightforward approach to generate the quantum anomalous Hall effect in topological insulator thin films. 

However, as far as we know, the magnetic impurity problems, i.e., the Kondo problems\cite{Hewsonbook}, in spin-orbit coupled systems are far from clear up to now. A clear understanding of the Kondo problem is essential to study the low temperature transport properties and the magnetic ordering problems in these systems. In this paper, we study the most fundamental quantity of the Kondo problems, the Kondo temperature, in two dimensional electron gas with Rashba spin-orbit coupling.  There are some previous works have studied the Kondo temperature of spin-orbit coupled systems\cite{PhysRevB.82.081309,PhysRevB.85.081107,PhysRevLett.108.046601,PhysRevB.84.193411,PhysRevB.84.041309}. A unitary transformation is applied to demonstrate\cite{PhysRevB.82.081309} that the spin-orbit coupling can be absorbed into the effective hopping amplitudes and it has no effect on the Kondo temperature. Mean-field calculations based on variational wave functions show\cite{PhysRevB.85.081107} that the Kondo temperature is exponentially altered by the spin-orbit coupling, because the density of state on the Fermi surface is changed by spin-orbit coupling. Perturbation renormalization group analysis based on the Schrieffer-Wolff transformation shows\cite{PhysRevLett.108.046601} that the Kondo temperature is exponentially enhanced by the Dzyaloshinskii-Moriya (DM) interaction between the local moment and the itinerant electrons. The numerical renormalization group (NRG) calculations show\cite{PhysRevB.84.193411} that the Kondo temperature is almost a linear function of Rashba spin-orbit coupling energy. A dependence of the Kondo temperature on the spin-orbit coupling is also expected for quantum dots\cite{PhysRevB.84.041309}, where due to additional parameters like left-right asymmetries in the tunneling couplings or magnetic fields more complicated behaviour including suppression may appear. These qualitatively different results show that the variation of Kondo temperature as a function of spin-orbit coupling in this system is still an open question. 

In this work, we use the Hirsch-Fye quantum Monte Carlo\cite{PhysRevLett.56.2521} (HFQMC) simulation to study the problem. HFQMC is a numerically exact method. It has been widely used to study the magnetic impurity problems in normal metals, i.e., the local moment of magnetic impurity\cite{PhysRevLett.56.2521,PhysRevB.38.433}, the Kondo effect\cite{PhysRevB.35.4901}, the Ruderman-Kittel-Kasuya-Yosida interaction between magnetic impurities\cite{PhysRevB.35.4943}, etc. Recently the HFQMC technique has been applied to study the local moment formation of Anderson impurities in dilute magnetic semiconductors\cite{PhysRevB.76.045220} and graphene\cite{PhysRevB.84.075414}, and the interaction between magnetic impurities in the bulk of topological insulators\cite{PhysRevB.89.115101}.

Let us present the model Hamiltonian here, the total Hamiltonian has three different terms: the bulk state of two-dimensional electron gas with Rashba spin-orbit coupling\cite{RolandWinkler2003,PhysRevB.84.193411,PhysRevLett.108.046601}, the magnetic impurity with strong on-site Coulomb interaction, and the hybridization between the electron gas and the impurity states, 
\begin{eqnarray}
&\mathcal{H}=\mathcal{H}_{\text{b}}+\mathcal{H}_{\text{d}}+\mathcal{H}_{\text{hyb}},\label{eq1}\\
&\mathcal{H}_{\text{b}}=\sum_{\bm{k},s=\uparrow,\downarrow}\psi_{\bm{k},s}^{\dagger}\left[\frac{\bm{k}^2}{2m}-\mu+\alpha(k_x\sigma_y-k_y\sigma_x)\right]\psi_{\bm{k},s},~~~\label{eq2}\\
&\mathcal{H}_{\text{d}}=\sum_{s=\uparrow,\downarrow}d_{s}^{\dagger}(\epsilon_d-\mu)d_{s}+Ud_{\uparrow}^{\dagger}d_{\uparrow}d_{\downarrow}^{\dagger}d_{\downarrow},\label{eq3}\\
&\mathcal{H}_{\text{hyb}}=\sum_{\bm{k},s=\uparrow,\downarrow}\left( V_{\bm{k}}\psi_{\bm{k},s}d_{s}+V_{\bm{k}}^{*}d^{\dagger}_{s}\psi_{\bm{k},s}\right),\label{eq4}
\end{eqnarray}
where $\psi_{\bm{k},s}^{\dagger}$ and $\psi_{\bm{k},s}$ are creation and annihilation operators of two-dimensional electron gas with momentum $\bm{k}$ (the Plank constant $\hbar$ has been set to be 1) and spin $s$. $m$ is the effective mass of two-dimensional electron gas. $\mu$ is the chemical potential. $\alpha$ is the strength of Rashba spin-orbit coupling. $\sigma_{x,y,z}$ are the spin Pauli matrices. $d_{\uparrow}^{\dagger}$ ($d_{\downarrow}^{\dagger}$) and $d_{\uparrow}$ ($d_{\downarrow}$) are creation and annihilation operators of spin-up (spin-down) impurity states, respectively. $\epsilon_d$ is the energy level of the impurity state. $U$ is the on-site Coulomb interaction. $V_{\bm{k}}$ is the hybridization between bulk electrons and impurity states. Without lose of generality, we set the hybridization to be short range coupled, so that the hybridization $V_{\bm{k}}$ is independent of momentum $\bm{k}$. 

Here we make a note on the units and parameters used in this paper. In the continuum limit, the summation over momentum is approximated by an integration, $\sum_{\bm{k}}=\frac{4\pi{N_0}}{k_T^2}\int_0^{\infty}\frac{d^2k}{(2\pi)^2}$, where $N_0$ is the number of sites, $k_T$ is determined by the band width $D=k_T^2/2m$, the total density of states (DOS) is $\rho=N_0/2D$, and converntionlly we define the hybridization $\Gamma=\pi\rho{V}^2$. We set $\Gamma=0.05\pi$ and $U=1.0$ in numerical calculation so that our results can be comparable with those given in numerical renormalization group (NRG) calculations. 

The bulk states can be integrated out to get the partition function of impurity states, 
\begin{eqnarray}
&\mathcal{Z}=\int\mathcal{D}d_s^{\dagger}\mathcal{D}d_s~e^{-\mathcal{S}}\label{eq5}\\
&\mathcal{S}=\int_0^{\beta}{\mathrm{d}\tau}\left\{d_{s}^{\dagger}\left[\frac{\partial}{\partial\tau}+\left(\epsilon_d-\mu\right)+\Sigma_d\right]d_s+Ud_{\uparrow}^{\dagger}d_{\uparrow}d_{\downarrow}^{\dagger}d_{\downarrow}\right\}~~\label{eq6}
\end{eqnarray}
where $\beta=1/{k_B}T$ is the inverse temperature, $k_B$ is the Boltzmann constant.  And $\Sigma_d$ is the self-energy of $d$-electrons, which is given in imaginary-frequency representation by, 
\begin{equation}
\Sigma_d(i\omega_n)=\frac{2\Gamma}{\pi}\left[\log\left(-\frac{i\omega_n+\mu}{4\pi}\right)+2{\zeta}\text{arctanh}{\zeta}+\gamma\right]\label{eq7}
\end{equation}
where $\zeta=\tilde{\alpha}/\sqrt{\tilde{\alpha}^2+2(i\omega_n+\mu)}$, $\tilde{\alpha}=\sqrt{m}\alpha$, and $\gamma=0.5772$ is the Euler-Mascheroni constant. from dimensional regularization(Make a note here: why I need the dimensional regularization). The DOS of bulk electrons is proportional to the imaginary part of $\Sigma_{d}(\omega+i\delta)$. 

\begin{figure}[t]
  \centering
  \includegraphics[width=0.8\columnwidth]{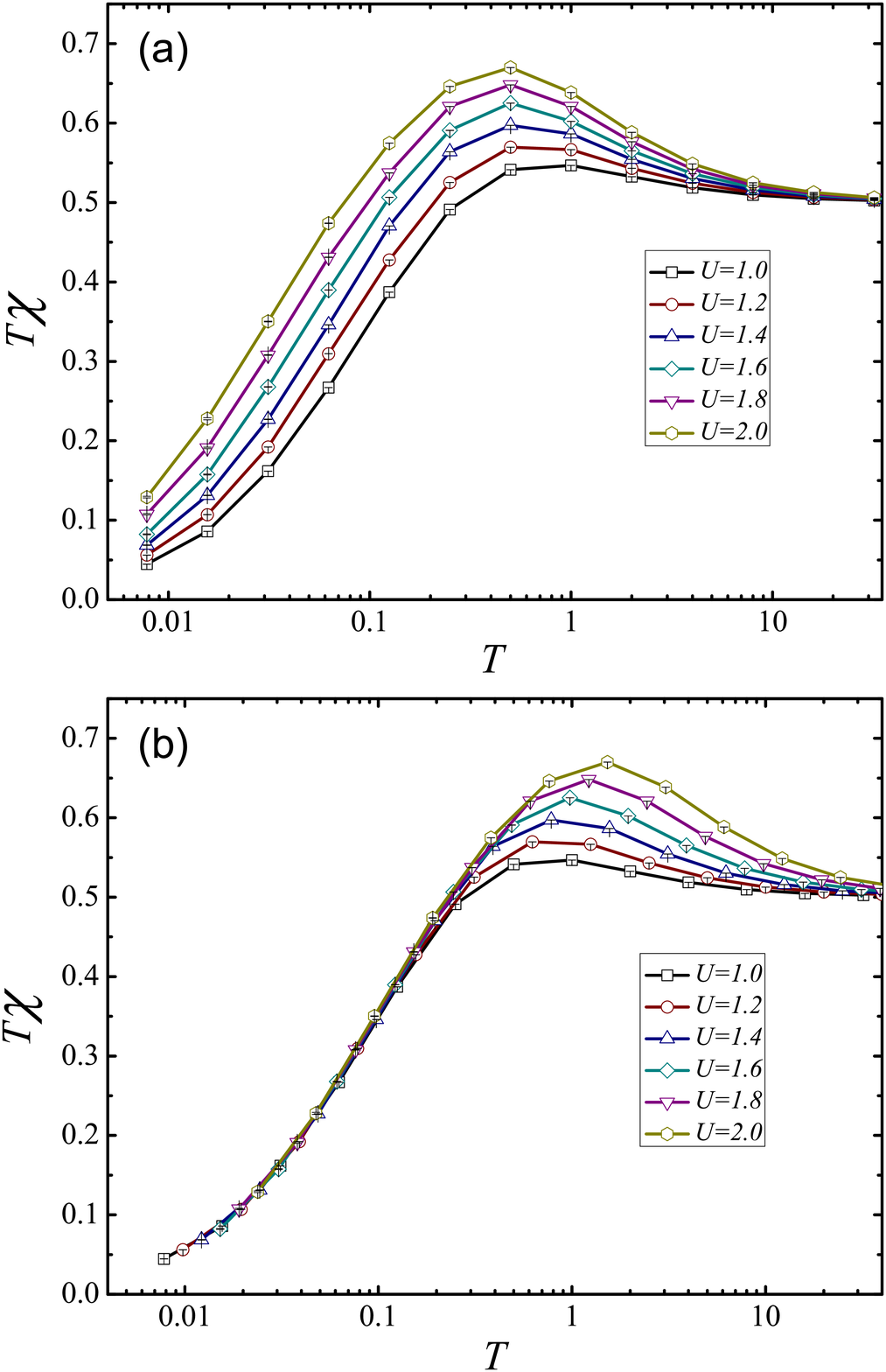}\\
  \caption{(a) The product of temperature $T$ and spin susceptibility $\chi$ as a function of temperature for different values of $U$. The spin-orbit coupling is chosen to be $E_{\alpha}=m\alpha^2/2=0.5$. Chemical potential $\mu=0.2$, and impurity energy $\epsilon_d-\mu=-U/2$. (b) The universal curves of $T\chi$ for rescaled temperatures (see the context for more details about rescaled temperatures).}\label{fig1}
\end{figure}

In HFQMC simulation, the Hubbard interaction is simulated by the statistical average over auxiliary Ising spin configurations. One can calculate many important static quantities, i.e., the local charge, the local moment, the spin susceptibility, etc., from the imaginary-time Green's function. The dynamical quantities, i.e., the spectral function, can be obtained in HFQMC by the maximal entropy method\cite{PhysRevB.41.2380}. The spin susceptibility $\chi$ is an important quantity to describe the Kondo problem, 
\begin{equation*}
\chi=\int_0^{\beta}d\tau\langle{S_z(\tau)S_z(0)}\rangle. 
\end{equation*}
where $S_z=d^{\dagger}_{\uparrow}d_{\uparrow}-d^{\dagger}_{\downarrow}d_{\downarrow}$. At very high temperature (i.e., $T\ge10$ in Fig. \ref{fig1}(a)), the total system behaves like free electron gas, which satisfies the Curie-Wiess law and $T\chi=1/2$\cite{PhysRevLett.56.2521}. At the intermediate region ($0.1<T<1$ in Fig. \ref{fig1}(a)), the impurity behaves as a local moment state. Our numerical simulation show that the square of the local moments $m^2$ for the parameters given in Fig. \ref{fig1} are enhanced in this intermediate region. At very low temperature, the local moment state is screened by the itinerant electrons, and the product of temperature and spin susceptibility satisfies the following universal relationship\cite{PhysRevB.21.1003}, 
\begin{equation}
\Phi\left(4T\chi(T)-1\right)=\ln\left(T/T_{K}\right), \label{eq8}
\end{equation}
where $\Phi$ is a universal function and $T_K$ is the Kondo temperature. We can use this relationship to fit the variation of the  Kondo temperature as a function of different parameters, i.e. Hubbard interaction, spin-orbit coupling, chemical potential, hybridization, etc.  

\begin{figure}[t]
  \centering
  \includegraphics[width=\columnwidth]{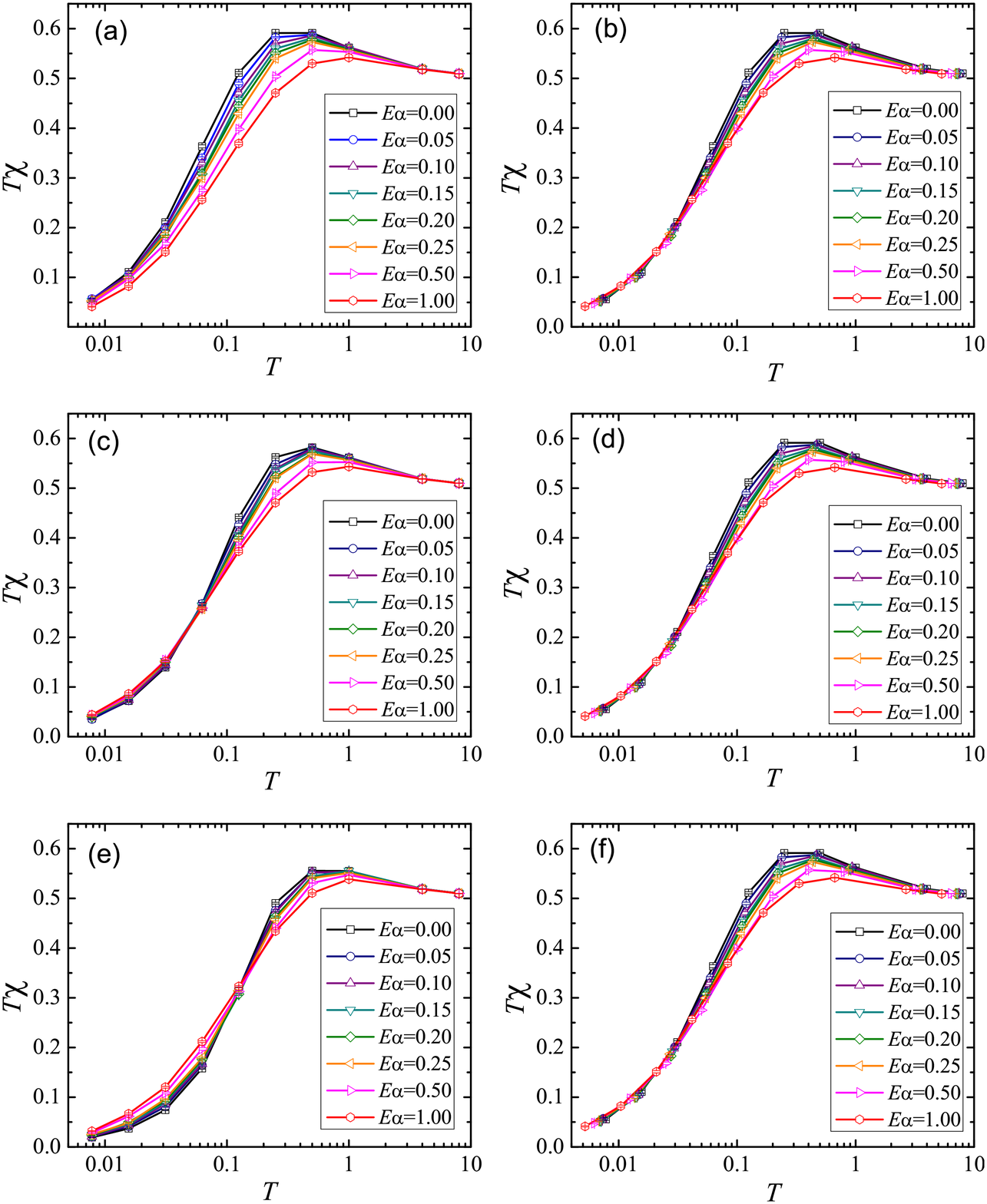}\\
  \caption{The product $T\chi$ for different impurity energies and different spin-orbit couplings. (a), (c) and (e) are the original data for different impurity energies $\varepsilon_d-\mu=-0.3$, $\varepsilon_d-\mu=-0.5$ and $\varepsilon_d-\mu=-0.7$ respectively. (b), (d) and (f) are the corresponding rescaled curves of (a), (c) and (e). Hubbard interaction $U=1.0$ and chemical potential $\mu=0.2$ are chosen. } \label{fig2}
\end{figure}

Before a more detailed discussion about the effect of spin-orbit coupling, we show the universal curves of spin susceptibility for different Hubbard interactions, shown in Fig. \ref{fig1}. One can see that the curves in Fig. \ref{fig1}(a) are almost equal-spacing under the logarithmic coordinate of temperature, which demonstrates that the Kondo temperature is exponentially altered by the Hubbard interaction. We can use the following formula to fit the Kondo temperature, 
\begin{equation}\label{eq9}
T_{K}(U)=T_{K}^{(0)}e^{-aU}.
\end{equation} 
Fig. \ref{fig1}(b) is the numerical fitting of the universal relationship Eq.(\ref{eq8}) about Fig. \ref{fig1}(a), where the horizontal coordinate is fixed and the longitudinal coordinate is rescaled for different Hubbard interactions. We find that the fitting result $a=1.11572$. This value is different from the one estimated from the "flat band" results 
\begin{equation}
T_K=\tilde{D}\sqrt{\rho{J}}\exp\left(-1/\rho{J}\right)\label{eq10}
\end{equation} 
where $\rho{J}=8\Gamma/\pi{U}$ and $\tilde{D}$ is the effective bandwidth. Eq. (\ref{eq10}) gives that $a=2.5$ for the hybridization $\Gamma=0.05\pi$ used here.  This result demonstrates that the density of states near the band edge is important to alter the Kondo temperature. 

\begin{figure}[t]
  \centering
  \includegraphics[width=0.8\columnwidth]{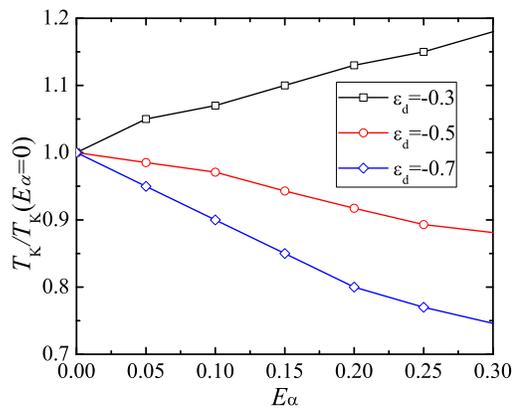}\\
  \caption{Kondo temperature as a function of spin-orbit coupling for different impurity energies. Hubbard interaction $U=1.0$ and chemical potential $\mu=0.2$. }
\label{fig3}
\end{figure}

\begin{figure}[t]
  \centering
  \includegraphics[width=0.47\columnwidth]{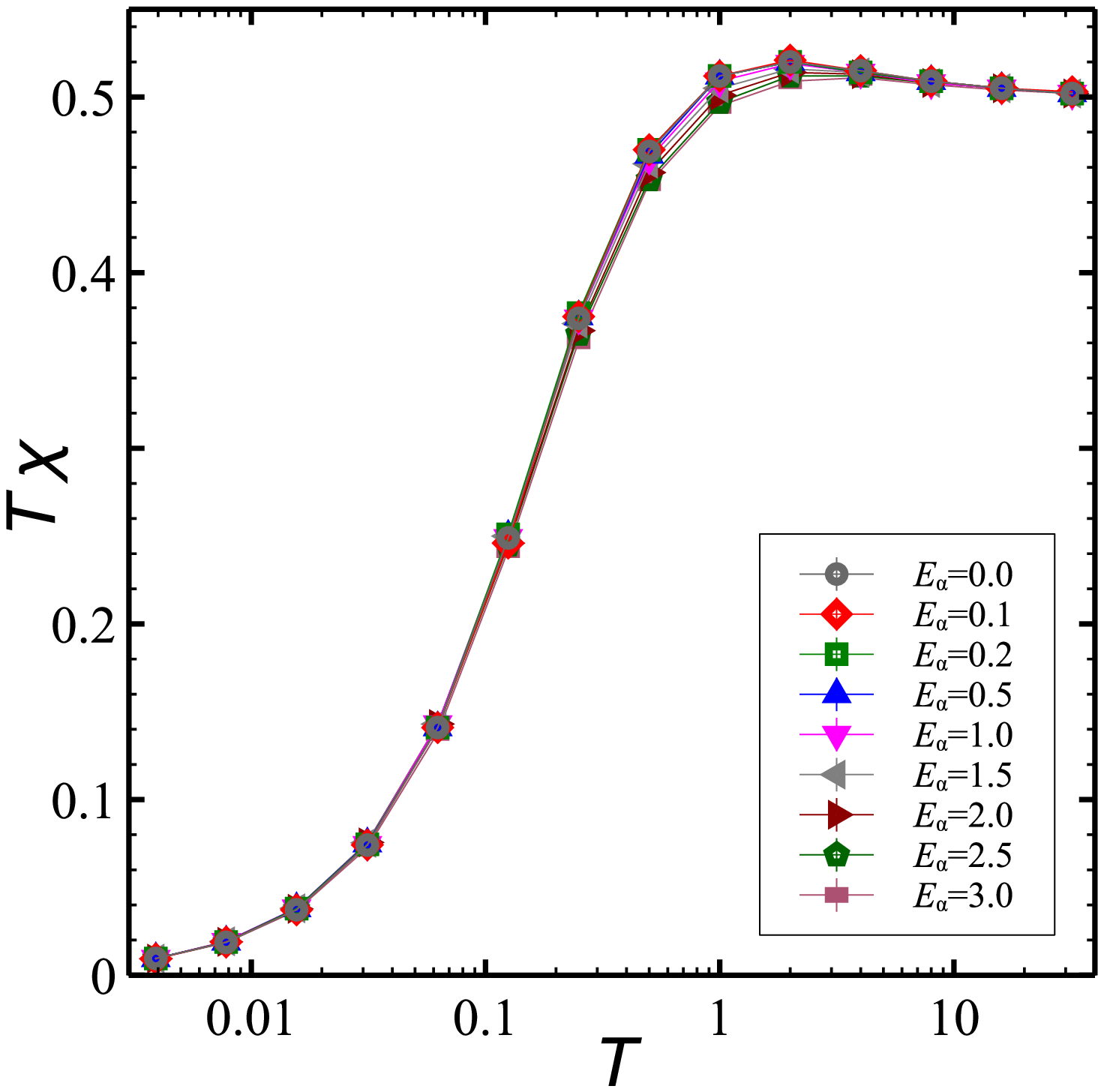}
  \includegraphics[width=0.47\columnwidth]{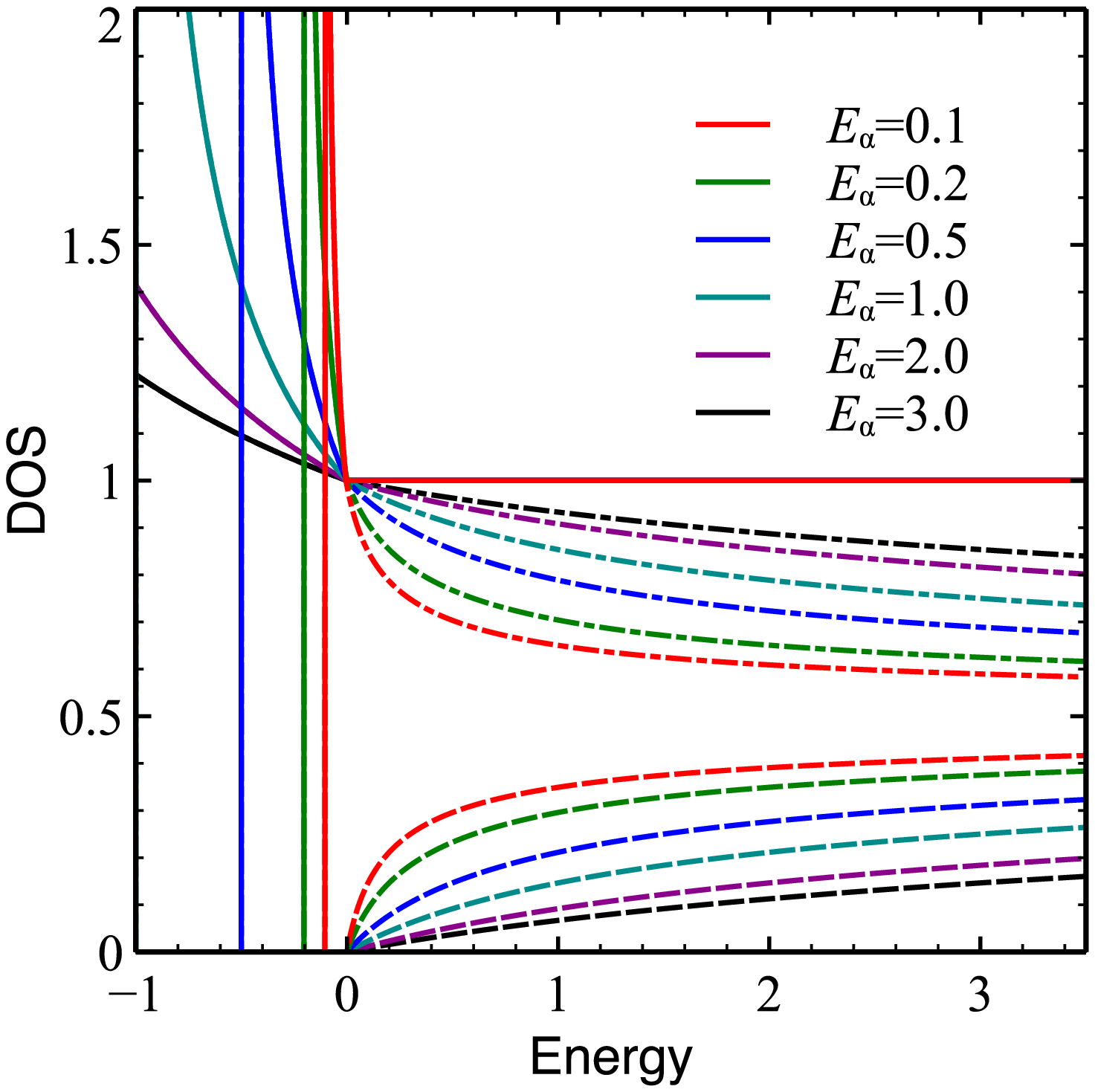}
  \caption{(a) The product $T\chi$ for different spin-orbit couplings when the chemical potential $\mu=3.0$, Hubbard interaction $U=1.0$ and impurity energy $\epsilon_d-\mu=-U/2$.  (b) The DOS of the bulk electrons for different spin-orbit couplings. The solid lines are the total DOS, dash lines and dotted-dash lines are the DOS of the two spitted bands.}
\label{fig4}
\end{figure}

Now we study the Kondo temperature for different spin-orbit couplings. There are three important approaches which may change the Kondo temperature: (1) The DM interaction between the local moment of the impurity and the itinerant electrons, which may expontionally enhance the Kondo temperature; (2) The spin-orbit coupling can spit the two-fold spin degeneracy of Fermi surface, so that the DOS on the two Fermi surface may change the Kondo temperature; And (3), although the total DOS on the Fermi surface is not altered by the spin-orbit coupling (see Fig. (\ref{fig4}) for more details), the divergence of DOS near the band edge may change the Kondo temperature. We find that our results support case (3). Now we show the numerical results. 

Fig. \ref{fig2} gives the product of temperature and spin susceptibility as a function of temperature for different spin-orbit couplings and different impurity energy levels. Figs. \ref{fig2}(a), \ref{fig2}(c) and \ref{fig2}(f) are original data for impurity energies $\epsilon_d-\mu=-0.3$, $\epsilon_d-\mu=-0.5$ and $\epsilon_d-\mu=-0.7$ respectively. Figs. \ref{fig2}(b), \ref{fig2}(d) and \ref{fig2}(f) are the rescaled curves. The numerical fitting of the Kondo temperatures are given in Fig. \ref{fig3}.  We find that our QMC results are qualitatively  consistent with those from NRG calculations. The Kondo temperatures are almost linear functions of the Rashba energy $E_{\alpha}=m\alpha^2/2$. However, there are some differences, (1) when the impurity energy $\epsilon-\mu=-0.5=-U/2$, the NRG calculations show that the Kondo temperature is enhanced by the spin-orbit coupling, our QMC simulations show that it is suppressed; (2) the effect of spin-orbit coupling in QMC simulation is much smaller than those given in NRG. These differences demonstrate that different identifications of the  Kondo temperature may have some quantitative differences. More importantly, one can find that, when the temperature is in the intermediate region ( i.e. $0.1<T< 1$), all of the three cases ($\epsilon_d-\mu=-0.3$, $-0.5$, $-.07$) have the tendency that the Kondo temperature is enhanced by the spin-orbit coupling. This demonstrates that maybe the perturbative renormalisation analysis is valid in this parameter region. However, when the temperature come into the Kondo region, i.e. $T<0.1$, the perturbative renormalisation analysis breaks down, and there may exist some crossover from the enhancement to the depression, i.e. $T\approx0.1$ in Fig. \ref{fig2}(e). 

These analysis show that the effect of DM interaction can be neglected at very low temperature. There are two other important effects which can change the Kondo temperature: the difference of the two Fermi surfaces and the divergence of the DOS near the band edge. Now we discuss the results when the chemical potential is far away from the band edge.  The numerical results are shown in Fig. \ref{fig4}, we calculate the product of temperature and the spin susceptibility as a function of temperature for various values of spin-orbit coupling from 0.1 to 3.0. In Fig. \ref{fig4}(b), we can see that the difference between the DOS of the two bands is dramatically enhanced. However, the universal curves in Fig. \ref{fig4}(a) are slightly changed and the Kondo temperature is almost unchanged by the spin-orbit coupling. 

So we conclude that the exponential enhancement of the Kondo temperature from perturbative renormalisation group analysis of the DM interaction between the impurity state and the bulk electrons breaks down in the low temperature Kondo region. Kondo temperature is almost a linear function of Rashba spin-orbit energy when the chemical potential is close to the edge of the conduction band, and when the chemical potential is far away from the band edge, the Kondo temperature is independent of the spin-orbit coupling. And the divergence of the DOS near the band edge is the most important factor to alter the Kondo temperature.

This work is supported by NSAF (Grant No. U1230202), Special Foundation for theoretical physics Research Program of China (Grant No. 11447167), and CAEP. 



\bibliographystyle{apsrev4-1}
\bibliography{ref}

\end{document}